\documentclass[a4paper, twoside, reqno, 12pt, dvips]{amsart}
\RequirePackage{fix-cm} 

\usepackage{fixltx2e}     

\usepackage{amssymb}
\usepackage{amsmath}
\usepackage{mathrsfs, dsfont}

\usepackage{a4wide}

\usepackage{acronym}

\usepackage{indentfirst}
\usepackage{graphicx}
\usepackage{psfrag}

\usepackage[usenames, dvipsnames, pdf]{pstricks}
\usepackage{epsfig}
\usepackage{pst-grad} 
\usepackage{pst-plot} 

\usepackage{subfigure}

\usepackage{longtable}

\usepackage[french,english]{babel}
\usepackage[latin1]{inputenc}

\usepackage{color}
\definecolor{oneblue}{rgb}{0,0.0,0.75}
\usepackage[colorlinks,
            urlcolor=oneblue,
            linkcolor=oneblue,
            citecolor=oneblue,
            bookmarksopen=false,
            pagebackref]{hyperref}

\numberwithin{equation}{section}

\renewcommand{\i}{\mathrm{i}}

\newcommand{\RE}{\mathrm{Re}}

\begin{document}

\title[The Force of a Tsunami on a Wave Energy Converter]{The Force of a Tsunami on a Wave Energy Converter}

\author[L. O'Brien]{Laura O'Brien}
\address{School of Mathematical Sciences, University College Dublin, Dublin, Ireland}
\email{loliwarm@gmail.com}

\author[P. Christodoulides]{Paul Christodoulides}
\address{Faculty of Engineering and Technology, Cyprus University of Technology, Limassol, Cyprus}
\email{Paul.Christodoulides@cut.ac.cy}

\author[E. Renzi]{Emiliano Renzi}
\address{School of Mathematical Sciences, University College Dublin, Belfield, Dublin 4, Ireland}
\email{Emiliano.Renzi@ucd.ie}

\author[D. Dutykh]{Denys Dutykh}
\address{School of Mathematical Sciences, University College Dublin, Belfield, Dublin 4, Ireland \and LAMA, UMR 5127 CNRS, Universit\'e de Savoie, Campus Scientifique, 73376 Le Bourget-du-Lac Cedex, France}
\email{Denys.Dutykh@univ-savoie.fr}
\urladdr{http://www.lama.univ-savoie.fr/~dutykh/}

\author[F. Dias]{Fr\'ed\'eric Dias$^*$}
\address{School of Mathematical Sciences, University College Dublin, Belfield, Dublin 4, Ireland \and CMLA, UMR 8536 CNRS, Ecole Normale Sup\'erieure de Cachan, Cachan, France}
\email{Frederic.Dias@ucd.ie}
\thanks{$^*$ Corresponding author}

\begin{abstract}
With an increasing emphasis on renewable energy resources, wave power technology is fast becoming a realistic solution. However, the recent tsunami in Japan was a harsh reminder of the ferocity of the ocean. It is known that tsunamis are nearly undetectable in the open ocean but as the wave approaches the shore its energy is compressed creating large destructive waves. The question posed here is whether a nearshore wave energy converter (WEC) could withstand the force of an incoming tsunami. The analytical 3D model of \textsc{Renzi} \& \textsc{Dias} (2012) \cite{Renzi2012a} developed within the framework of a linear theory and applied to an array of fixed plates is used. The time derivative of the velocity potential allows the hydrodynamic force to be calculated.
\end{abstract}


\maketitle

\tableofcontents

\section{Introduction}

\begin{figure}[htbp]
\begin{center}
\includegraphics[width=0.8\textwidth]{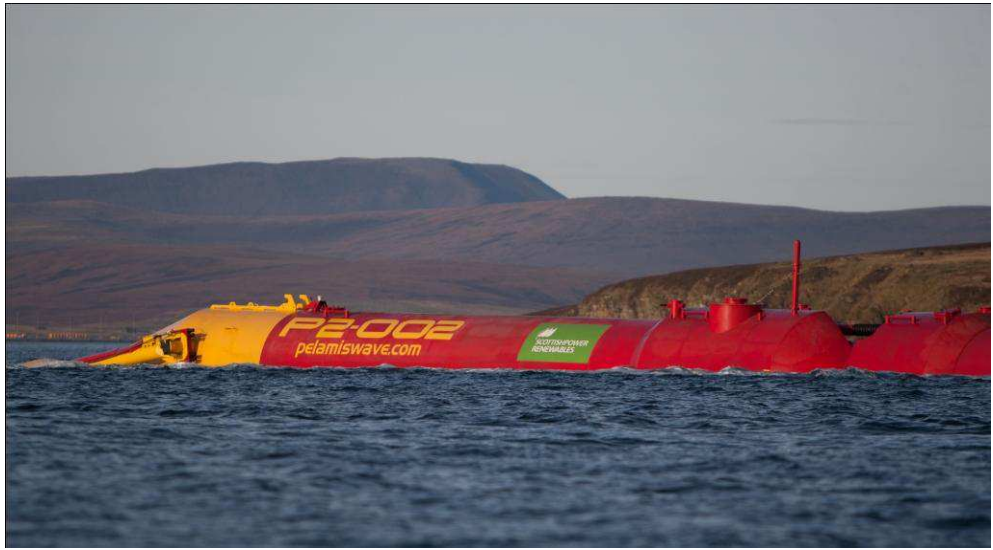}
\caption{Pelamis Wave Power Device, picture from \url{http://www.pelamiswave.com}}
\label{Pelamis}
\end{center}
\end{figure}

\begin{figure}[htbp]
\begin{center}
\includegraphics[width=0.8\textwidth]{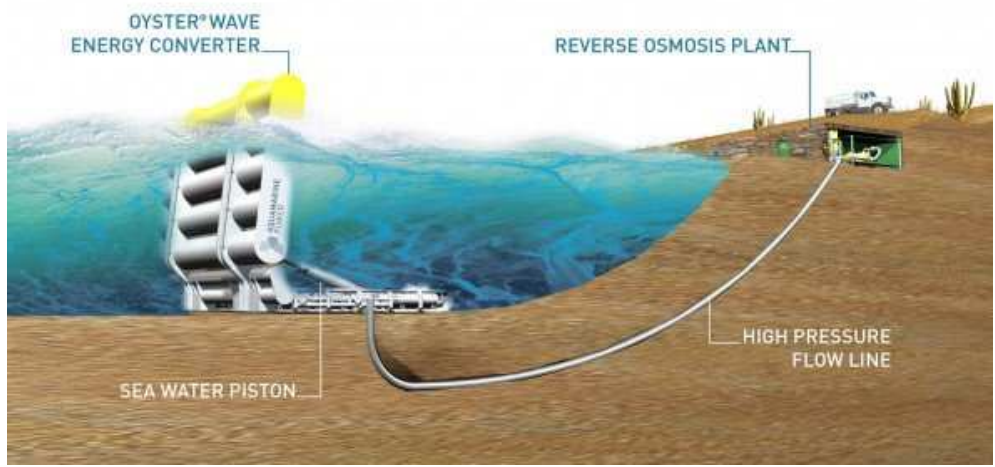}
\caption{Oyster Wave Power Device picture from \url{http://www.aquamarinepower.com}}
\label{oyster}
\end{center}
\end{figure}

Wave energy devices are slowly becoming a reality. Various prototypes are now being tested in harsh sea conditions (storms). What about tsunamis? Even if offshore wind turbines seem to have survived the 2011 tsunami in Japan, it is legitimate to ask whether WECs will resist tsunamis. For deep sea WECs, such as Pelamis (Figure \ref{Pelamis}), tsunamis are not anticipated to be a threat since they are located far from the shore (the present Pelamis prototype operating at EMEC, Orkney, is located $2$ km from the shore). On the other hand, for nearshore WECs, such as Oyster (Figure \ref{oyster}) it is important to take a closer look at the effect of tsunamis (the present Oyster prototype operating at EMEC, Orkney, is located $500$ m from the shore). Unfortunately there is very few tsunami wave data away from the shoreline. One exception is the Mercator yacht, anchored $1.6$ km away from the shore in Thailand during the 2004 Indian Ocean tsunami. The water depth was about $12-13$ m and the yacht experienced four major waves, one ``depression" wave ($2.8$ m) and three ``elevation" waves ($3.8$ m, $1.7$ m and $4.2$ m) \cite{Rossetto2011}. And the problem is quite different from the problem of wave forces acting on flap-type storm surge barriers \cite{Tomita2003} because the periods involved are different.

\textsc{Kajiura} (1977) \cite{Kajiura1977} considers the amplification of tsunamis which advance toward shore over a gentle slope using Green's law for tsunamis,
\begin{equation}
\frac{A_1}{A_2} = \left( \frac{\lambda_2}{\lambda_1} \right)^{\frac{1}{2}} =  \left( \frac{h_2}{h_1} \right)^{\frac{1}{4}} .
\label{greens}
\end{equation}
where $A_i$ and $\lambda_i$ are the amplitude and wavelength of a tsunami at a depth of $h_i$, at two different positions $i = 1,2$. Four dimensionless parameters are defined in Table \ref{dim_less}. These parameters are used to compare the importance of linear, non linear and dispersive effects.

\begin{table}
\begin{center}
\begin{tabular}{|l|l|l|l|}
\hline
Relative & Wave & Wave & Ursell \\ 
Height & Shallowness & Steepness & Number \\
\hline
&&&\\
$\displaystyle{\epsilon_i = \frac{A_i}{h_i}}$ & $ \displaystyle{\delta_i = \frac{h_i}{\lambda_i}}$ & $\displaystyle{\gamma_i = \frac{A_i}{\lambda_i}}$ & $\displaystyle{U_{ri} = \frac{\epsilon_i}{\delta_{i}^2}}$\\
\hline
\end{tabular}
\end{center}
\vspace{0.5em}
\caption{Dimensionless numbers}
\label{dim_less}
\end{table}%

Green's Law (\ref{greens}) implies that for a tsunami at 2 different positions $1$ and $2$,
\begin{eqnarray}
\epsilon_2 &=& \epsilon_1 \left(\frac{h_1}{h_2}\right)^{5/4} \;\;\;\;\;\;\;\;\; \delta_2 = \delta_1 \left(\frac{h_1}{h_2}\right)^{-1/2} \nonumber \\
\gamma_2 &=& \gamma_1 \left(\frac{h_1}{h_2}\right)^{3/4} \;\;\;\;\;\;\;\;\; U_{r2} = U_{r1} \left(\frac{h_1}{h_2}\right)^{9/4}.
\label{dim_less_1_2}
\end{eqnarray}

If one takes a typical tsunami wave with $A_1 = 1$ m, $h_1 = 3$ km, $\lambda_1 = 100$ km then the corresponding dimensionless parameters arising from Table \ref{dim_less} are shown in column 1 of Table \ref{tsu_vals}.

\begin{table}
\begin{center}
\begin{tabular}{|l|l|l|}
\hline
&&\\
 & $h_1 = 3$ km & $h_2 = 31$ m\\
\hline
&&\\
$\epsilon_i$ & $3.3$x$10^{-4}$ & $10^{-1}$\\ 
&&\\
$\delta_i$ & $0.03$ & $3$x$10^{-3}$\\
&&\\
$\gamma_i$ & $10^{-5}$ & $3$x$10^{-4}$\\
&&\\
$U_{ri}$ & $0.367$ & $1.08$x$10^4$\\
&&\\
\hline
\end{tabular}
\end{center}
\vspace{0.5em}
\caption{Values of dimensionless numbers for tsunami at two positions $i = 1,2$ according to Equations \ref{dim_less_1_2}.}
\label{tsu_vals}
\end{table}%

These values indicate that linear theory  can be used to describe the behaviour of the wave up to a certain depth and a slight dispersive effect should be included for the wave travelling over very large distances. However, when considering small travel distances over a few kilometers the linear shallow water equations are sufficient. As the wave approaches the shore, finite amplitude (non linear) effects come into play when $\epsilon_2 \approx 10^{-1}$. According to Equations (\ref{dim_less_1_2}) this occurs at a depth of $h_2 = 31$ m. Assuming a sea bed slope of $0.02$ this occurs at a distance of approximately $1.5$ km from the shore which is about one seventh of the wavelength of a tsunami with a period of $10$ minutes. The dimensionless parameters corresponding to this depth are shown in column 2 of Table \ref{tsu_vals}. The wave steepness is $\gamma_2 \approx 0.0003$ and the Ursell number is $U_{r2} \approx 10^4 \gg 1$, indicating that dispersion is relatively minor compared with the non-linearity except for the front part of the wave. From these considerations, it is reasonable to conclude that at this distance from the shore there is a shift in importance from linear to non linear effects. Therefore linear shallow-water equations used offshore should be matched to the inner solution of the nonlinear shallow-water equations at a distance from shore of about a seventh of a wavelength of the tsunami. As a first approximation, linear theory is used here to predict the force exerted on a WSC.

\section{Model description}

We consider here the following idealized problem: a flap-type structure mounted at the sea bottom pierces the surface of the ocean. The structure is assumed to be fixed. What is the load on the flap due to a tsunami wave?

Authorities tend to classify the different forces acting on a structure due to a tsunami in the following way. In the document entitled ``Development of design guidelines for structures that serve as tsunami vertical evacuation sites'' \cite{Yeh2005} several forces are described by a number of design codes \cite{FEMA2000, DPP2000, ASCE2010} which include loading requirements based on equations given by \cite{Dames&Moore1980}:
\begin{itemize}
\item Hydrostatic Forces: Occur when standing or slowly moving water encounters a structure. They are caused by an imbalance of pressure due to a differential water depth on opposite sides of structure and act perpendicular to the surface.
\item Buoyancy Force: Concerns structures with little resistance to lift eg. light wood frame buildings, basements, or swimming pools. These act vertically through the center of mass of the displaced volume.
\item Hydrodynamic Force: Caused by water flowing at a moderate to high velocity around a structure. These are a combination of the lateral forces caused by the pressure forces from the moving mass of water and the friction forces generated as the water flows around the structure. They include frontal impact, drag along the sides, and suction on the downstream side. This force is a function of flow velocity, fluid density and structural geometry. 
\item Breaking Wave Force: This force is taken as the hydrodynamic force if the wave breaks on the structure. When considering a breaking wave, generally the two structures of interest are piles/columns and walls. Waves that break obliquely incident to the wall (not perpendicular) result in a lower force. The net force is assumed to act at the still water elevation. It is also assumed that a breaking wave against a wall causes a reflected or standing wave and the crest of the wave is some height above the still water elevation. 
\item Surge Force: Another variety of hydrodynamic force caused by the leading edge of a surge of a tsunami impinging on a structure. 
\item Impact Force: Results from debris or any object transported by floodwaters, striking against a structure.
\end{itemize}

Assuming that the load is mainly hydrodynamic, even within this idealized framework it is not clear what the main force is going to be. The suggested loading for a solid wall facing the shoreline by \cite{Yeh2005} (ignoring impact forces) is given by one of three forces: a breaking wave force 
\begin{equation}
F_{brkw} = (1.1C_p+ \varsigma)\rho g d_s^2 w,
\end{equation} 
a surge force
\begin{equation}
F_s = 4.5 \rho g h^2,
\end{equation}
or a hydrodynamic force
\begin{equation}
F_d = \frac{1}{2} \rho C_d A_r u_p^2,
\end{equation}
where $\rho$ is the water density, $g$ is the acceleration due to gravity, $C_{p} \in [1.6, 3.5]$ is the dynamic pressure coefficient, $\varsigma = 1.9$ or $2.4$ is a hydrostatic coefficient, $w$ is the width of the wall, $A_r$ is the area of the wall, $h$ is the surge height, $C_d\approx 1.5$ is the drag coefficient, $u_p = 2\sqrt{g d_s}$ is the design flood velocity and $d_s$ is the surge depth.

Furthermore, \textsc{Chen} \& \textsc{Scawthorn} (2003) \cite{Chen2003} give the form proposed by \textsc{Cross} (1967) \cite{Cross1967} for the force on a seawall, of width $w$, 
\begin{equation*}
F_{wall} = \frac{1}{2} \rho g w \eta^2(x_{wall},t) + C_f(t) \rho w \eta(x_{wall},t) v^2,
\end{equation*}
where $\eta(x_{wall},t)$ is the water surface elevation on the wall, $v $ is the surge or bore velocity, $w$ is the width of the wall and $C_f = (1 + \tan\theta^{1.2})$ where $\tan\theta$ is the slope of the front face of the bore as it impacts the wall which is estimated using experimental and theoretical data.

These formulas produce values that are highly dependent on coefficient estimates and on the design flood velocity, which is highly conservative $u_p = 2\sqrt{gd_s}$, twice that of the wave speed of a tsunami given by shallow water theory. Another way (a more precise way) to look at forces is through the integral of the stress tensor. Since viscous effects are neglected, the only contribution comes from the pressure term. In turn the pressure term can be evaluated through Bernoulli's equation. In the linear model $p = -\rho(gz + \Phi_t)$, where $-\rho g z$ is the hydrostatic pressure and $-\rho \Phi_t$ is simply the dynamic pressure. In the fully nonlinear model $p = -\rho(gz + \Phi_t + \frac{1}{2}\left|\nabla\Phi\right|^2)$, so the dynamic pressure has an additional term $-\rho \frac{1}{2}\left|\nabla\Phi\right|^2$. We focus here on the linear model and calculate the dynamic pressure difference across a fixed plate.

\section{Linear model}

\begin{figure}
  \centerline{\includegraphics[width=0.7\textwidth]{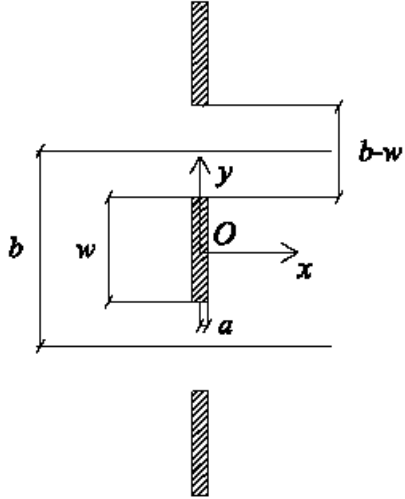}}
   \caption{Geometry of the array of WECs (plan view).}
\label{fig:plate_geom}
\end{figure}

Consider an infinite array of equally spaced thin plates in the open ocean used for the purpose of wave energy conversion (e.\ g.\ WECs). The analysis of the scattering problem, in which the flaps are held fixed in incoming waves, is used here to calculate the velocity potential and so, the pressure exerted on the system. This is important in order to investigate whether an array of nearshore WECs would survive the impact from a tsunami. Periodicity of the problem allows the geometry to be reduced to that of a single plate within two waveguides at a mutual distance $b$, as shown in Figure \ref{fig:plate_geom}.

With reference to Figure \ref{fig:plate_geom}, the plate is represented by a rectangular box of width $w$ and thickness $2a$, hinged along a straight foundation at a distance $c$ from the bottom of the ocean of depth $h$. The plate is in the middle of a channel of total width $b$. A plane reference system of coordinates $\boldsymbol{x}=(x,y,z)$ is also set, with $x$ on the centre line of the channel, $y$ along the axis of the plate and $z$ positive upwards. Monochromatic waves of frequency $\omega$ are incoming from the left with wave crests parallel to the plate.

\subsection{Model equations}

The theoretical basis of the mathematical model is provided by \cite{Renzi2012a} and summarised here. Within the framework of a linear potential-flow theory, the velocity potential $\Phi(x,y,z,t)$ must satisfy the Laplace equation
\begin{equation}
\nabla^2\Phi(x,y,z,t) =0 \label{eq:lapl}
\end{equation}
in the fluid domain, with $\nabla$ the nabla operator. On the free-surface, the kinematic-dynamic boundary condition
\begin{equation*}
\frac{\partial^2\Phi}{\partial t^2}+g\frac{\partial \Phi}{\partial z}=0, \quad z=0
\end{equation*}
is applied, with $g$ the acceleration due to gravity. Absence of normal flux at the bottom and through the lateral walls of the channel requires
\begin{equation}
\frac{\partial\Phi}{\partial z}=0, \quad z=-h; \quad \frac{\partial\Phi}{\partial y}=0,\quad y=\pm b/2 \label{eq:bottom}
\end{equation}
respectively.

A no-flux boundary condition must be applied on the lateral surfaces of the plate, yielding
\begin{equation}
\frac{\partial\Phi}{\partial x}  =  0, \quad x=\pm a,\; |y|<w/2.
\label{eq:bcflap2lin}
\end{equation}

Since the total thickness of the plate $2a\ll b$, the thin-plate approximation can be made \cite{Linton2001} by which the boundary condition on the plate (\ref{eq:bcflap2lin}) is restated at $x=\pm 0$. Finally, the reflected and transmitted wave field respectively on the weather side and the lee side of the plate must be both outgoing at large distances from the origin.

\subsection{Solution}

The system of governing equations (\ref{eq:lapl})--(\ref{eq:bcflap2lin}) is solved via the introduction of a complex spatial potential $\phi(x,y,z)$ such that 
\begin{equation}
\Phi=\RE\left\lbrace\phi(x,y,z)e^{-i\omega t}\right\rbrace. 
\label{eq:spatpot}
\end{equation} 

Due to the linearity of the problem, the spatial potential $\phi$ is analysed by the classical decomposition\begin{equation}
\phi=\phi^I+\phi^D,
\label{eq:phisdecomp}
\end{equation}
 where
\begin{equation}
\phi^I(x,y,z) = -\frac{\i A_0 g}{\omega\cosh kh}\cosh k(z+h) e^{-\i kx} 
\label{eq:incidentwav}
\end{equation}
is the potential of the incident wave and $\phi^D$ the potential of the diffracted waves (see \cite{Linton2001, Mei2005}). In Equation (\ref{eq:incidentwav}) $A_0$ and $k$ are respectively the wave amplitude and wavenumber, the latter depending on the wave frequency according to the dispersion relation $\omega^2 = gk\tanh kh$.

Application of the Green integral theorem to the governing system of equations (\ref{eq:lapl})--(\ref{eq:incidentwav}) yields an integral equation for $\phi^D$ with a strong kernel singularity. The latter is solved with a new analytical method based on the careful treatment of the singularity (for details see \cite{Renzi2012a}).

The solution is expressed in terms of a series of Chebyshev polynomials of the second kind $U_p$:
\begin{multline}
\phi(x,y,z) = \phi^I(x,y,z) 
 -\frac{1}{4\sqrt{2}}\, \RE\Bigg\{\i gA_0 w\,kx 
 \frac{\cosh k(z+h)}{\left(gh+(g/\omega)^{2} \sinh^2 kh\right)^{1/2}}\times \\ \sum_{p=0}^{N}\beta_{p}\sum_{m=-\infty}^{+\infty}\int_{-1}^{1} \left(1-u^2\right)^{1/2} U_{p}(u) \frac{H_1^{(1)}\left(k\sqrt{x^2+(y-\frac{1}{2} wu-m)^2} \right)}{\sqrt{x^2+(y-\frac{1}{2} wu-m)^2}}\, du\Bigg\}. \label{eq:pot}
\end{multline}
In the latter expression $H_1^{(1)}$ the Hankel function of first kind and first order. Finally, the $\beta_p$, $p=0\dots N\in\mathbb{N}$, are the complex solutions of a system of linear equations which ensures that the boundary condition on the plate (\ref{eq:bcflap2lin}) is respected. This system is solved numerically with a collocation scheme, therefore the solution (\ref{eq:pot}) is partly numerical.

Once the potential is known, the dynamic pressure on the plate can be found 
\begin{equation}
  p = -\rho\frac{\partial\Phi}{\partial t} = \rho \, \RE\left\{\i\omega\phi e^{-\i\omega t}\right\}.
\label{eq:pressure}
\end{equation}

This is directly related to the hydrodynamic force $F$, acting on the plate by integrating over the wet surface of the body $S_B$, i.\ e.\
\begin{equation}
  F = \iint_{S_B} p(x,y,z,t) \,dA,
\label{eq:FDe}
\end{equation} 
where $dA=dx dy$ the infinitesimal area on the plate.

Substituting Equation (\ref{eq:pot}) into (\ref{eq:pressure}) at $x = 0^\pm$, the pressure jump across the plate is given by
\begin{equation}
\Delta p = p(0^-, y, z, t) - p(0^+, y, z, t)
= \rho\RE\{\i\omega [ \phi^D(0^-, y, z) - \phi^D(0^+, y, z) ] e^{-\i \omega t}\}
\label{eq:pressjump}
\end{equation}
since
\begin{equation}
\phi^I(0^-, y, z) - \phi^I(0^+, y, z) = 0. 
\label{eq:phi_I_jump}
\end{equation}
Therefore, the maximum pressure jump can be found
\begin{equation}
|\Delta p| = \rho  g \omega A_0(1- u^2) \sum_{p=0}^{N}\beta_p U_p(u)  \frac{\sqrt{2} \cosh k(z+h)}{\left(h+\omega^{-2} \sinh^2 kh\right)^{1/2}},
\label{eq:abs_press_jump}
\end{equation}
where $u = 2y/w$. Note that Equation (\ref{eq:abs_press_jump}) is truncated at $p=N$ making it an approximate expression, which converges as $N \rightarrow \infty$.

\section{Results}

\begin{figure}
  \centering
  \subfigure[Tsunami]{%
  \includegraphics[width=0.48\textwidth]{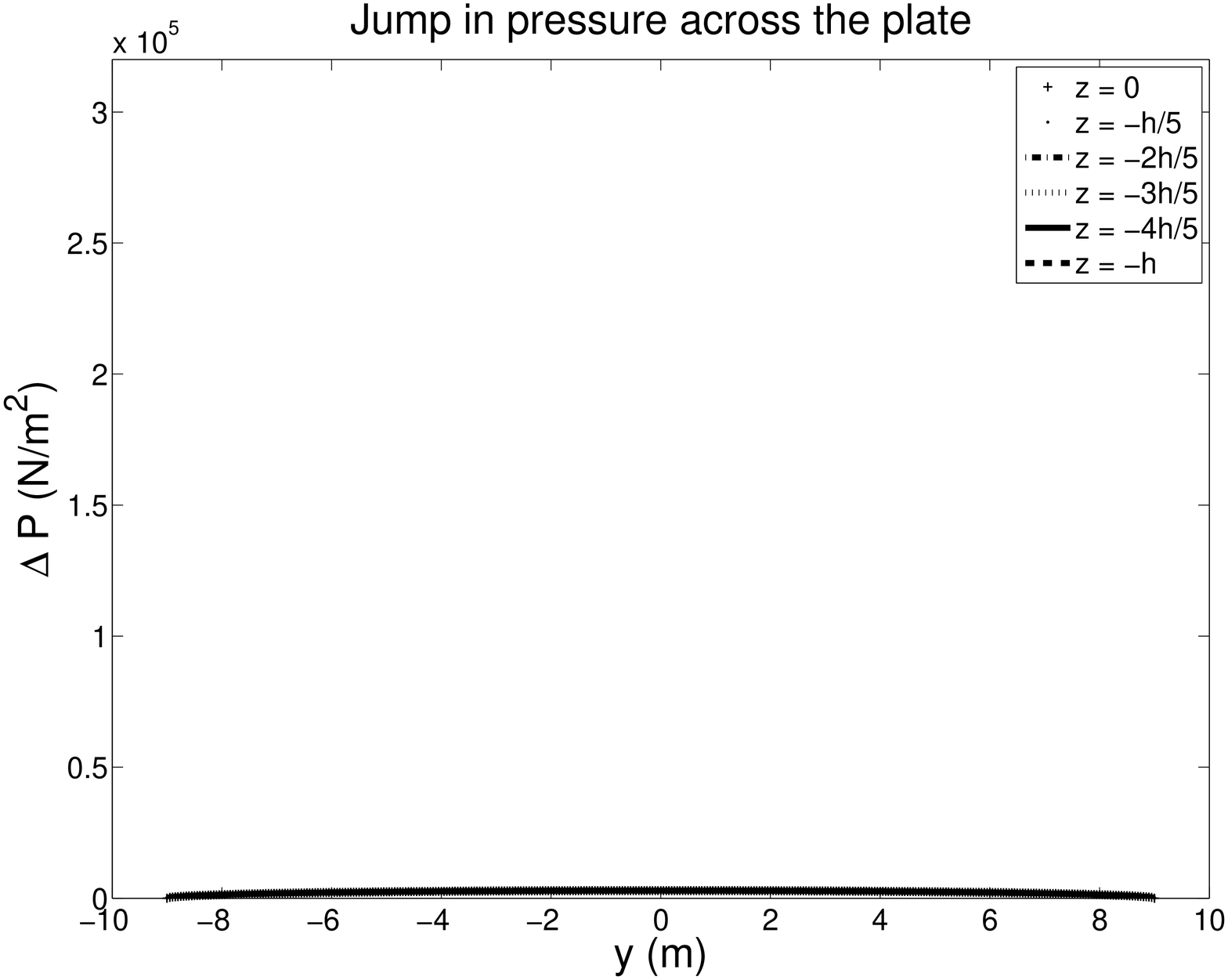}}
  \subfigure[Swell]{%
  \includegraphics[width=0.48\textwidth]{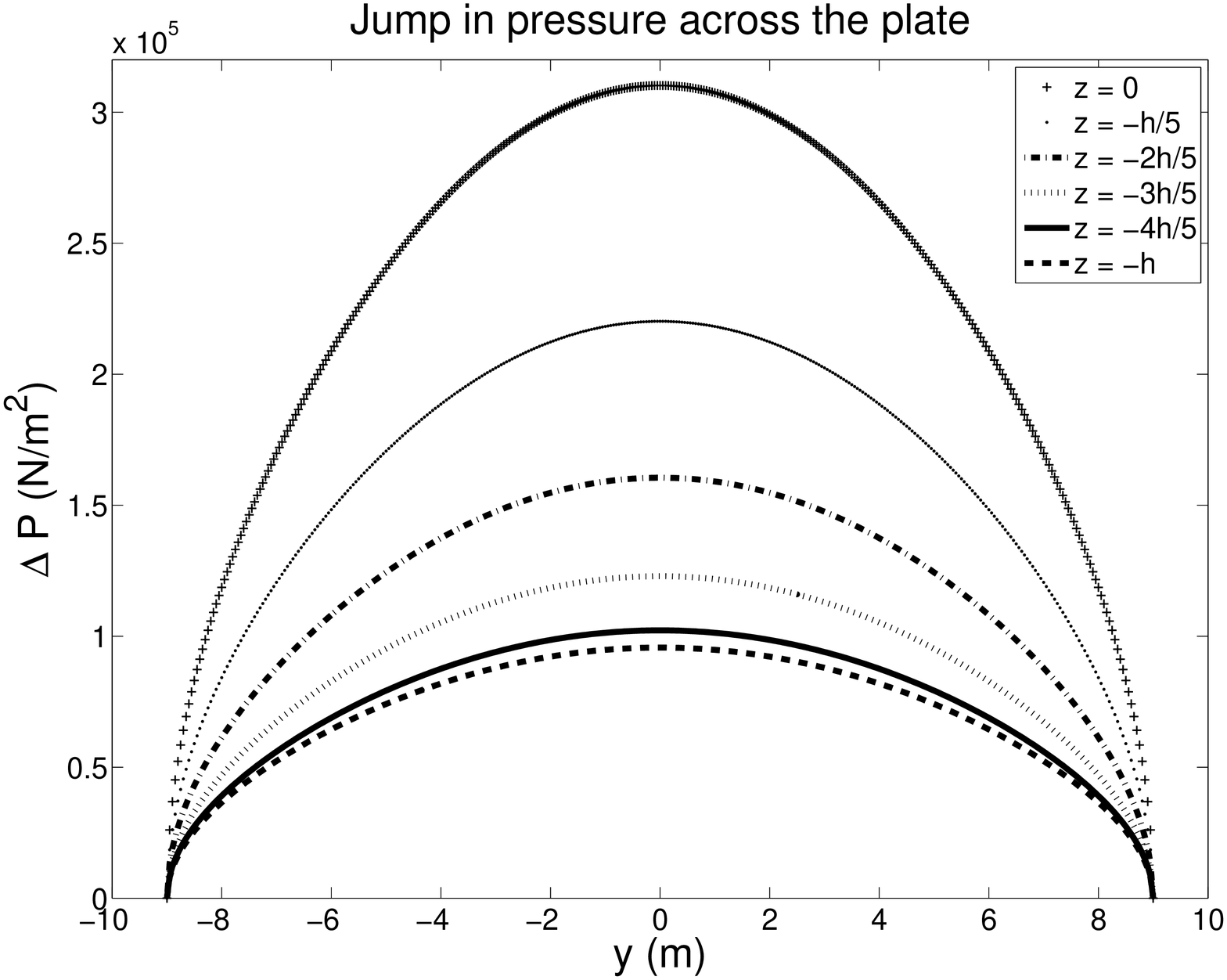}}
   \caption{Various jumps in pressure across an 18 m plate for a typical tsunami (a) and a typical swell (b), in a depth of $h = 10.9$ m at 6 depths from the free surface to the ocean floor.}
  \label{fig:pressure_jump}
\end{figure}

The force of a tsunami on an array of nearshore WECs at a depth of $10.9$ m is analysed. If the tsunami has an amplitude of $1$ m offshore at a depth of $3$ km, then according to Green's law for tsunamis (\ref{dim_less_1_2}) the amplitude of the wave will be approximately $4$ m when it hits the devices. Applying the linear model from \cite{Renzi2012a} we approximate the WECs as an array of a fixed plates with a spatial period $b = 91.6$ m and width $w = 18$ m and determine the pressure exerted on one plate from a tsunami with amplitude $4$ m and period $10$ minutes. The pressure jump across the plate is shown in Figure \ref{fig:pressure_jump} (a). It is plotted against $y$ which runs along the axis of the plate and calculated at $6$ equally spaced depths from the still water level to the sea floor. The greatest overall pressure difference is felt at the center of the plate ($y = 0$) and is zero at the edges of the plate ($y = \pm 9$ m) but is invariant with depth. The maximum value is $\Delta p \approx 3$x$10^3$ N/m$^2$ $ = 0.03$ bar.

In order to compare these results to a standard sea state, the pressure jump exerted by a typical swell with amplitude $3$ m and period $5$ s impacting on the plate is shown in Figure \ref{fig:pressure_jump} (b). This clearly shows how the pressure changes with depth, the maximum effect felt at the free surface and decreasing towards the sea floor. Also the magnitude is much greater than that from the tsunami, with a maximum $\Delta p \approx 3$x$10^5$ N/m$^2$ = $3$ bar. From these results we can conclude that the tsunami load exerted on the plate does not vary with depth since it is such a long wave relative to the depth. Moreover, the magnitude of load exerted by the tsunami is approximately $100$ times less than that of a normal swell. We can therefore assume that an array of nearshore WECs would easily withstand the force from a tsunami according to linear theory. However, as previously noted, non linear effects will start to become important at approximately $1.5$ km from the shore so further research into the non linear effects on the plate needs to be done.

\section{Conclusions}

\begin{figure}[htbp]
\begin{center}
\subfigure[]{%
\includegraphics[width=0.48\textwidth]{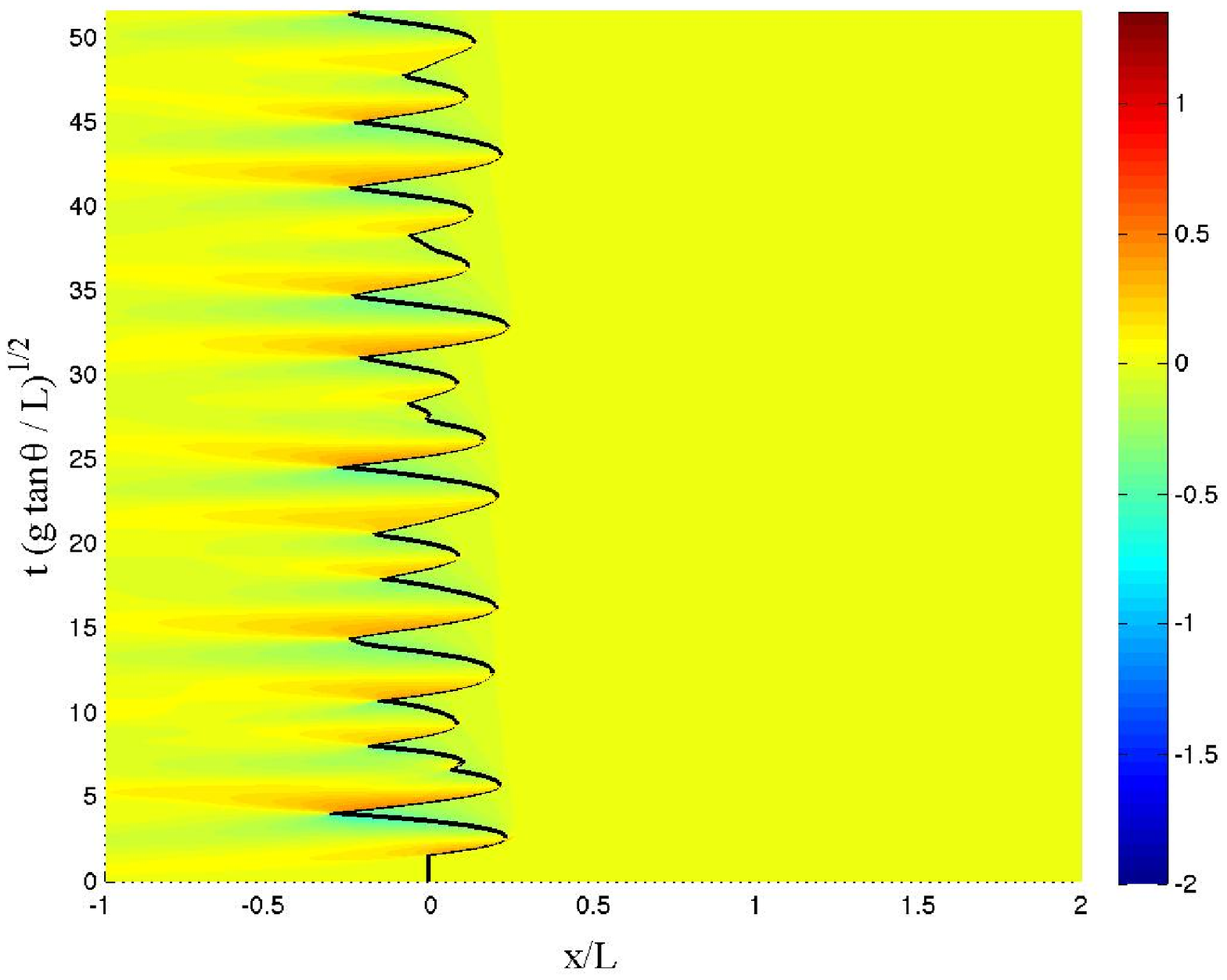}}
\subfigure[]{%
\includegraphics[width=0.48\textwidth]{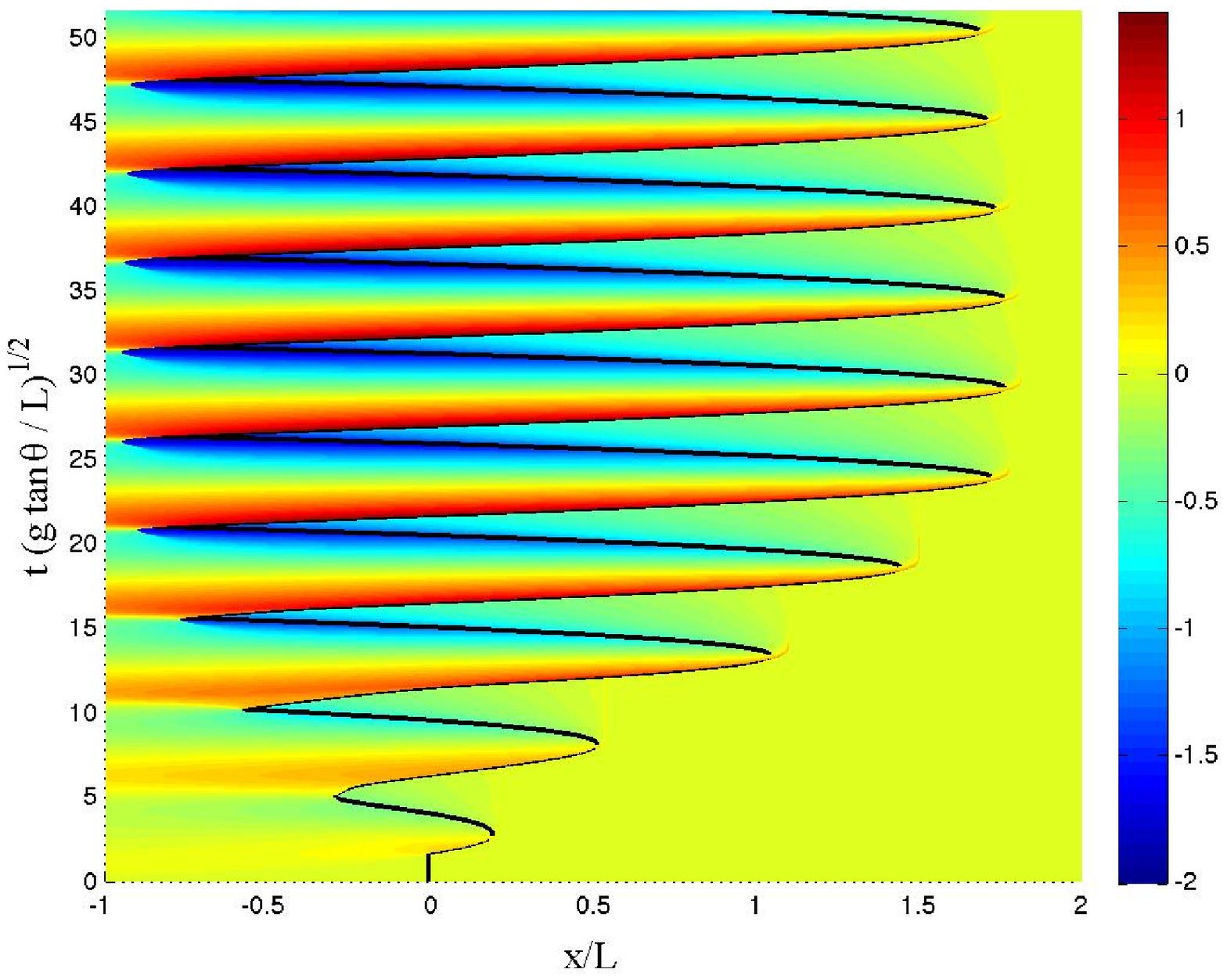}}
\end{center}
\caption{(a) Non resonant and (b) resonant non dimensional velocities from a monochromatic wave at $x = L = 12.5$ m on a sloping beach with slope $\tan\theta = 0.13$ and initial shoreline at $x = 0$ \cite{Stefanakis2012}.}
\label{fig:vel_themis}
\end{figure}

\begin{figure}[htbp]
\begin{center}
\subfigure[Bathymetry of a fixed plate on a sloping sea bed.]{
\label{Bathy}
\includegraphics[width=0.36\textwidth]{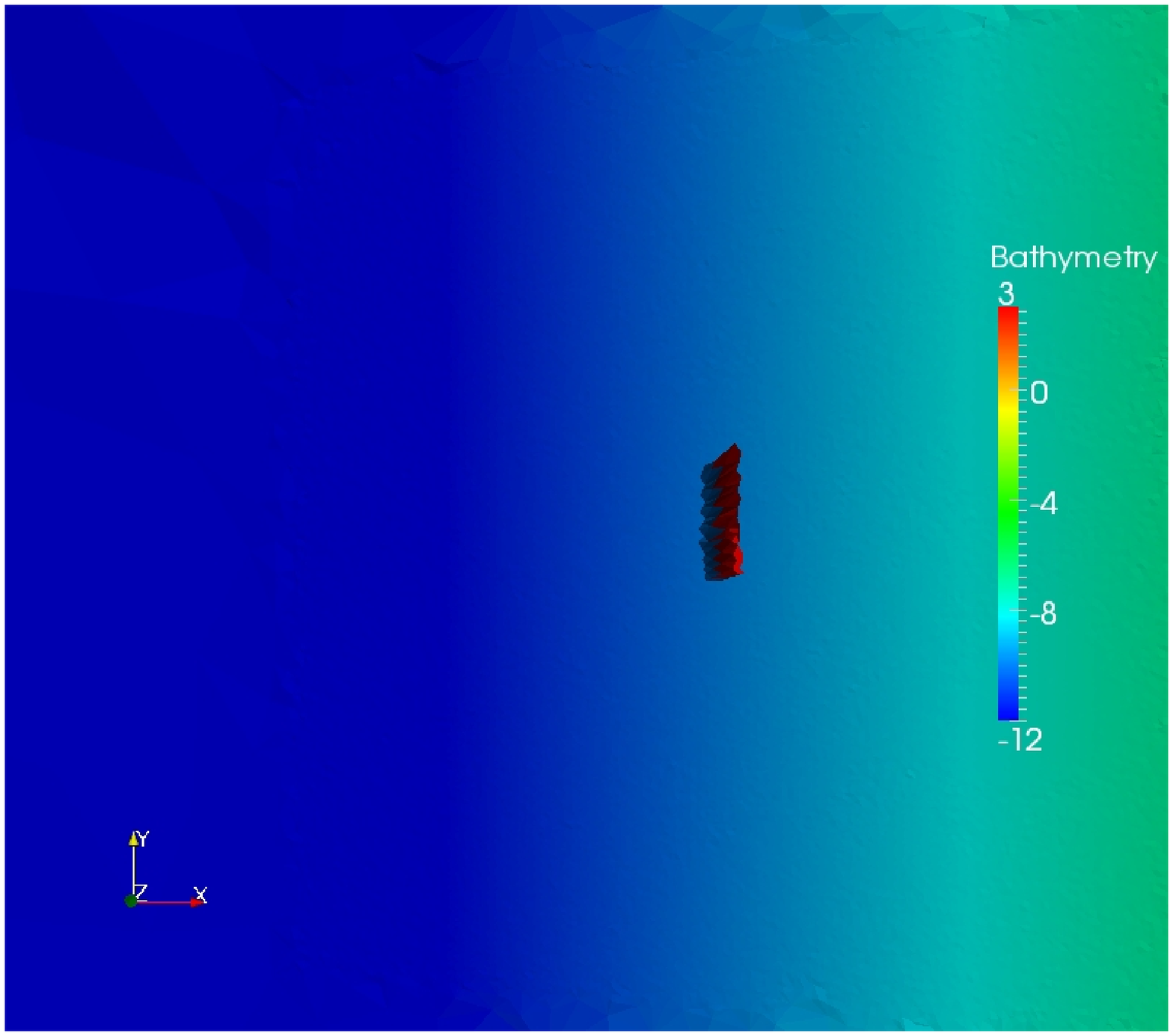}
}
\subfigure[Free surface $\eta$ after a tsunami has inundated the shore and it begins to recede.]{
\label{1575}
\includegraphics[width=0.47\textwidth]{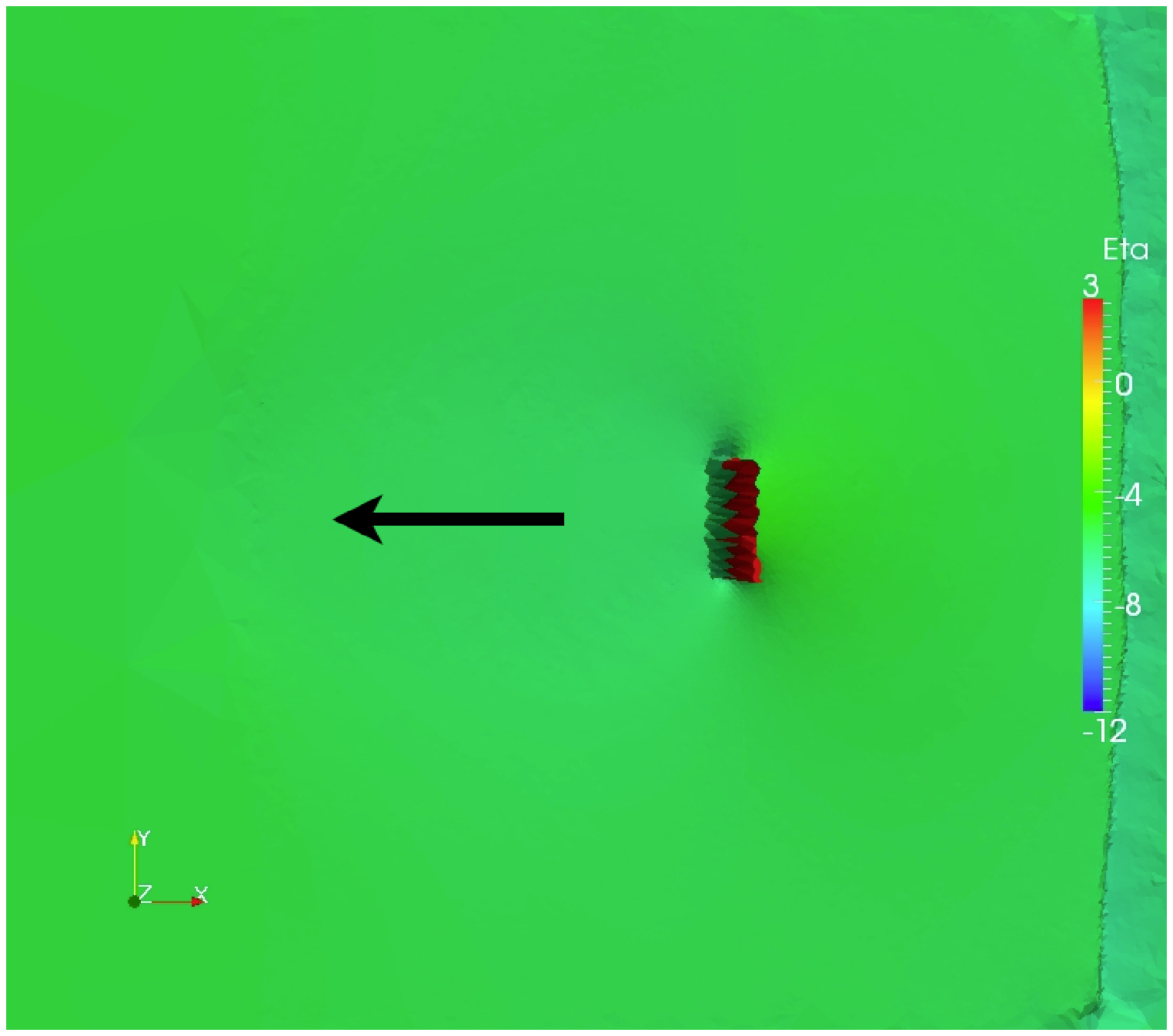} 
}
\subfigure[$45$ seconds later, the plate is left on dry land.]{
\label{1625}
\includegraphics[width=0.47\textwidth]{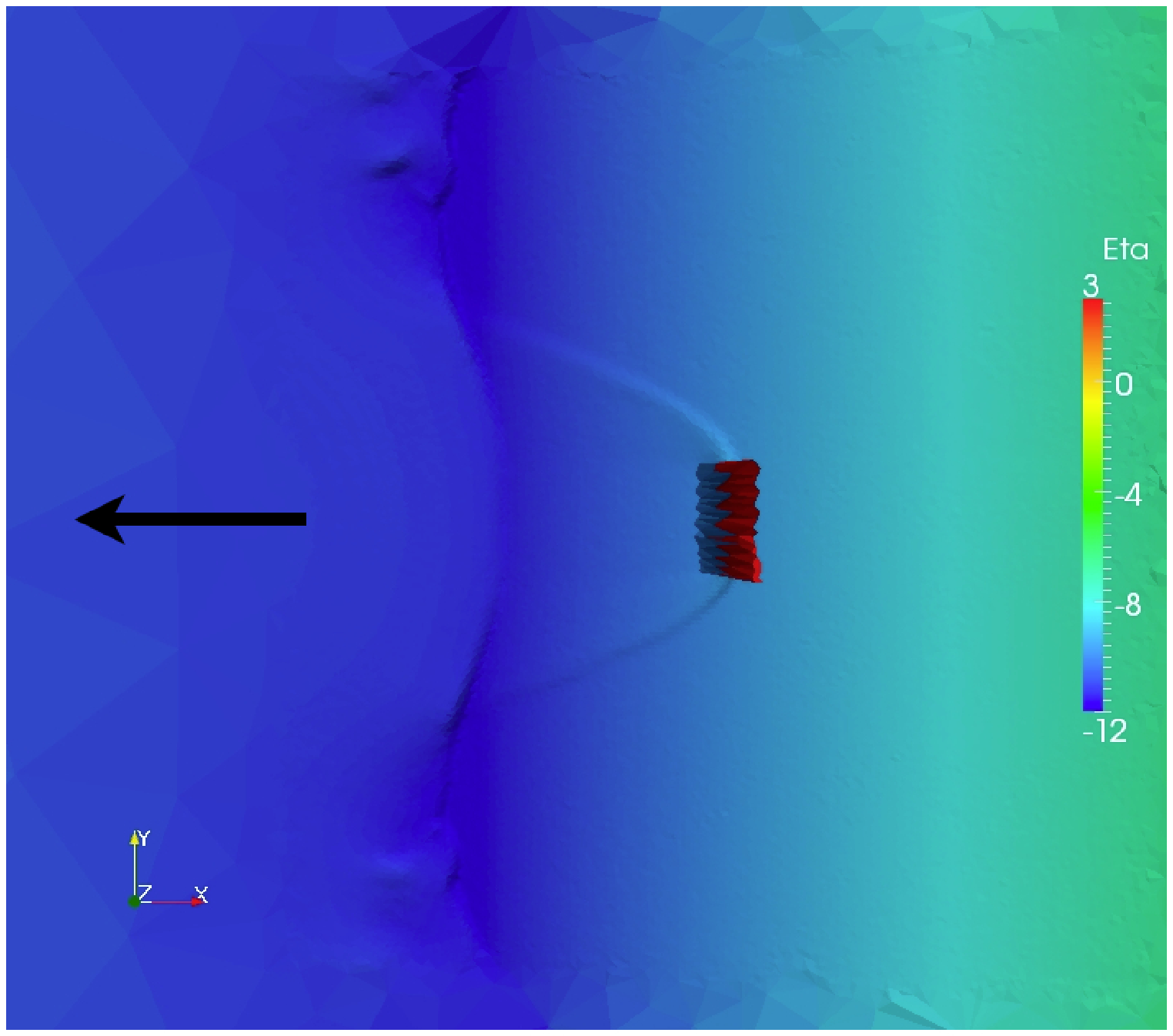} 
}
\subfigure[Another $45$ seconds later, a subsequent wave impacts the plate.]{
\label{1675}
\includegraphics[width=0.47\textwidth]{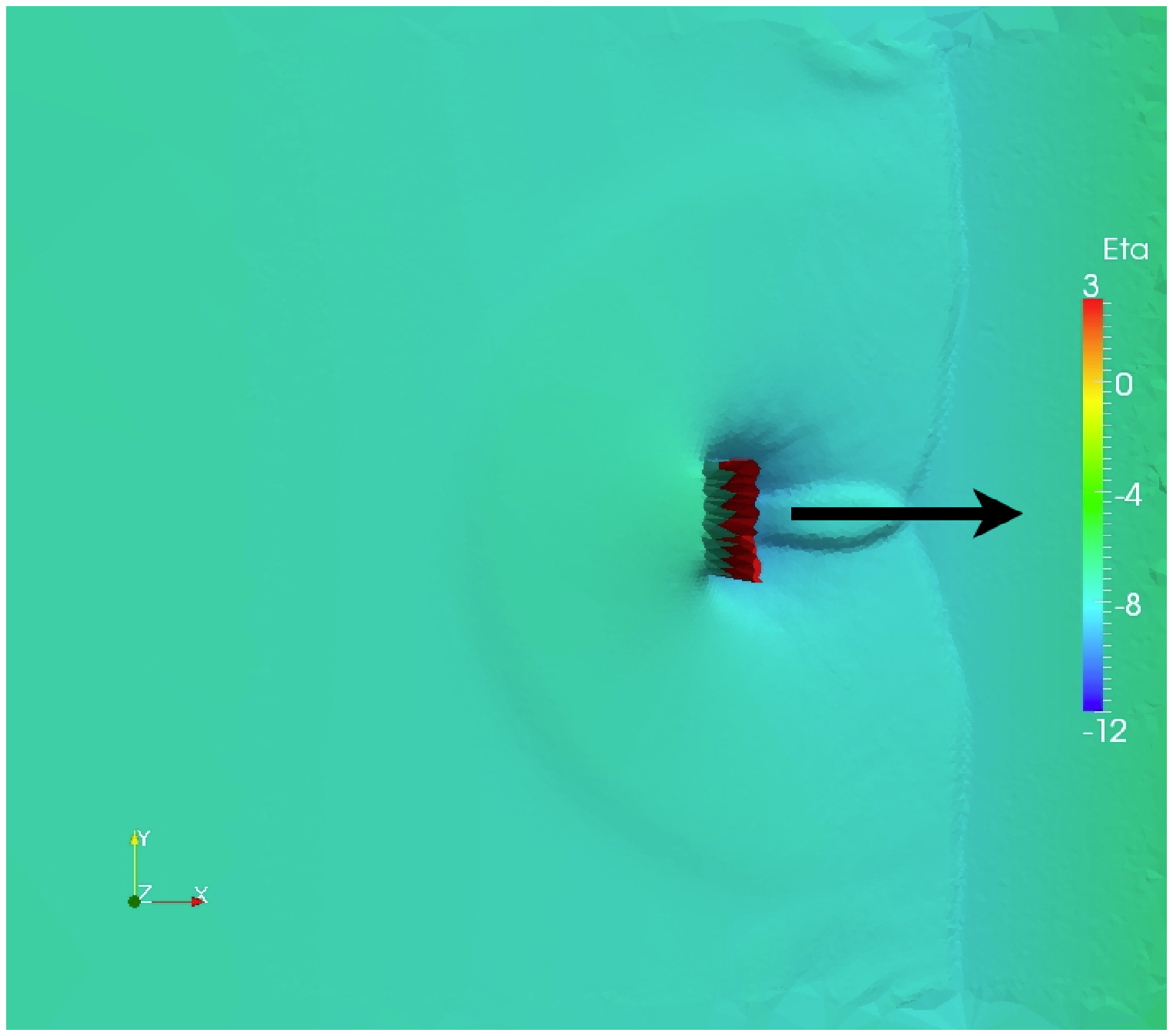} 
}
\end{center}
\caption{VOLNA simulation from multiple tsunami waves hitting a $13$ m high plate in an initial depth of $10$m on a sea bed with slope $0.03$ (plane view).}
\label{fig:Volna_sim}
\end{figure}

The hydrodynamic load of a tsunami on an array of nearshore WECs was investigated here. First, the difficulty in knowing whether to use linear or non linear theory was demonstrated, showing that WECs of this type are usually located close to the boundary of dominance between linear and non linear effects. Different forces suggested by standard tsunami design codes were reviewed displaying the variety of formulas and their reliance on estimated coefficients and a conservative velocity estimate. Applying the linear model from \cite{Renzi2012a} to an array of fixed plates, a first approximation for the hydrodynamic loading on a WEC was calculated through determining the jump in the $-\rho \Phi_t$ term. Results showed that the loading for a typical tsunami was invariant with depth and maximum loading is felt at the center of the plate. By comparison with the loading from a typical swell, it was shown that the maximum force of a tsunami on a nearshore WEC will be approximately one hundreth of the magnitude of a regular sea state. We therefore conclude that an array of WECs will withstand a tsunami. However, further research needs to be done on the non linear effects on nearshore WECs, in particular the effects of a sloping sea bed and multiple waves. \textsc{Stefanakis} \emph{et al}. (2011) \cite{Stefanakis2011} demonstrated resonant phenomena between the incident wavelength and the beach slope within the framework of the nonlinear shallow water equations in one dimension for multiple tsunami waves. A comparison between the velocities of resonant and non resonant states from \cite{Stefanakis2012} are shown in Figure \ref{fig:vel_themis}. Furthermore, if after the first wave recedes the device is left on dry land, a second wave may act as a shock on the plate and do more damage than it would to a partially submerged device. This effect is demonstrated using a two dimensional non linear shallow water solver, VOLNA \cite{Dutykh2009a} in Figure \ref{fig:Volna_sim} (note the wave is not exactly symmetric due to an unstructured triangular mesh being employed). We believe that dangerous configurations could be found with more detailed investigations.

\section*{Acknowledgements}

The authors would like to acknowledge the support provided by the Science Foundation Ireland (SFI) under the project {\it High-end computational modelling for wave energy systems}, by the Framework Program for Research, Technological Development, and Innovation of the Cyprus Research Promotion Foundation under the Project A$\Sigma$TI/0308(BE)/05, by the Irish Research Council for Science Engineering and Technology (IRCSET) and by Aquamarine Power.

\bibliography{biblio}
\bibliographystyle{plain}

\end{document}